\documentclass[aps,prl,reprint,superscriptaddress, longbibliography]{revtex4-1}
\usepackage{graphicx}
\usepackage{graphics}
\usepackage{epsfig}
\usepackage{epstopdf}
\usepackage{placeins}
\usepackage{lipsum}

\usepackage{array} 
\usepackage{multirow}

\usepackage{xcolor}
\usepackage[colorlinks,linkcolor=blue,citecolor=blue,urlcolor=blue] {hyperref}
\usepackage{amsmath}

\newcommand{\DDS}{$\mathrm{Dy_2ScN@C_{80}}$}
\newcommand{\DSD}{$\mathrm{Dy_2S@C_{82}}$}

\newcommand{\TTS}{$\mathrm{Tb_2ScN@C_{80}}$}

\newcommand{\DDO}{$\mathrm{Dy_2O@C_{82}}$}
\newcommand{\Deff}{$\Delta_\mathrm{eff}$}

\begin{document}


\title{Quantum tunneling of the magnetization in systems with anisotropic 4f ion pairs: Rates from low temperature zero field relaxation}

\author{Thomas Greber}
\email{greber@physik.uzh.ch}
\affiliation{Physik-Institut, Universit\"at Z\"urich, Winterthurerstrasse 190, CH-8057 Z\"urich, Switzerland}


\date{\today}

\begin{abstract}
Anisotropic open shell 4f ions have magnetic moments that can be read and written as atomic bits.
If it comes to qbits where the phase of the wave function has to be written, controlled and read, it is of advantage to rely on more than one atom that carries the quantum information of the system because states with different susceptibilities may be addressed. 
Such systems are realized for pairs of lanthanides in single molecule magnets, where four pseudospin states are found and mixed in quantum tunneling processes.
For the case of endohedral fullerenes like \DSD\ or \TTS\ the quantum tunneling of the magnetisation is imprinted in the magnetisation lifetimes at sub-Kelvin temperatures. A Hamiltonian that includes quantum tunneling of the magnetisation predicts the lifting of the zero field ground state degeneracy and non-linear coupling to magnetic fields in such systems. 

\end{abstract}

\maketitle

Since the description of the covalent molecular bond in the hydrogen molecule by Heitler and London \cite{Heitler1927} the concept of hybridisation established as a central pilar of quantum mechanics. 
Hybridisation also applies for state separation of ammonia NH$_3$ ground states out of a superposition of the two possibilities to arrange the nitrogen atom above or below the H$_3$ plane, which lead to the demonstration of stimulated microwave emission and manifested coherent light matter interaction \cite{Zeiger1954}. These concepts  apply as well for magnetic moments or spins that can be changed upon level crossing in a coherent fashion \cite{Khomitsky2022}.
While the quantum tunneling of the magnetisation survives as well in mesoscopic systems like crystals of single molecule magnets \cite{Thomas1996}, eventually it is desirable to maintain and control coherence in single single molecule magnets possibly with electromagnetic radiation.

Single molecule magnets (SMM's) feature bistable spin configurations with lifetimes in the order of seconds or longer \cite{Gatteschi2006} and the search for the best systems proceeds via the investigation of ensembles of SMM's.

Single lanthanide double decker molecules realized first mononuclear complexes with SMM behaviour where a 4f$^{\,8}$ Tb$^{3+}$ ion played the role of the bistable magnet \cite{Ishikawa2003}. Single ion SMM's were further developed where today's molecules reach hystereses above liquid nitrogen temperature \cite{Layfield2018}. 
Besides this search for highest blocking temperatures single single ion SMM's were used for read out of nuclear spin states \cite{Wernsdorfer2012} and implementation of quantum algorithms \cite{Godfrin2017}.
In parallel spin control and stability on single lanthanide ions on surfaces was obtained \cite{Natterer2017,Forrester2019}.

For the roadmap to single atom quantum devices it is important to explore the controlled arrangement and interference of more than one open shell atom.
In this aspect endofullerenes provide the unique opportunity of arranging up to three lanthanides in otherwise impossible very close distance ($<\,$0.4~nm) inside a magnetically quiet carbon shell \cite{Stevenson1999}.


The ground states of axially anisotropic 4f ion pairs can be described with a pseudospin model, where the magnetic moments on the two ions may assume two  orientations \cite{Westerstrom2014}. 
The ligand fields lift the Hund degeneracies of the 4f ions \cite{Rinehart2011}.
If they are dominated by anions like S$^{2\text{-}}$ or N$^{3\text{-}}$, cations like Tb$^{3+}$, Dy$^{3+}$ or Ho$^{3+}$ are expected to display axial anisotropy and pointing towards the anion or away from it.

\begin{figure}[b]
    \centering
    \includegraphics[width=7.5cm]{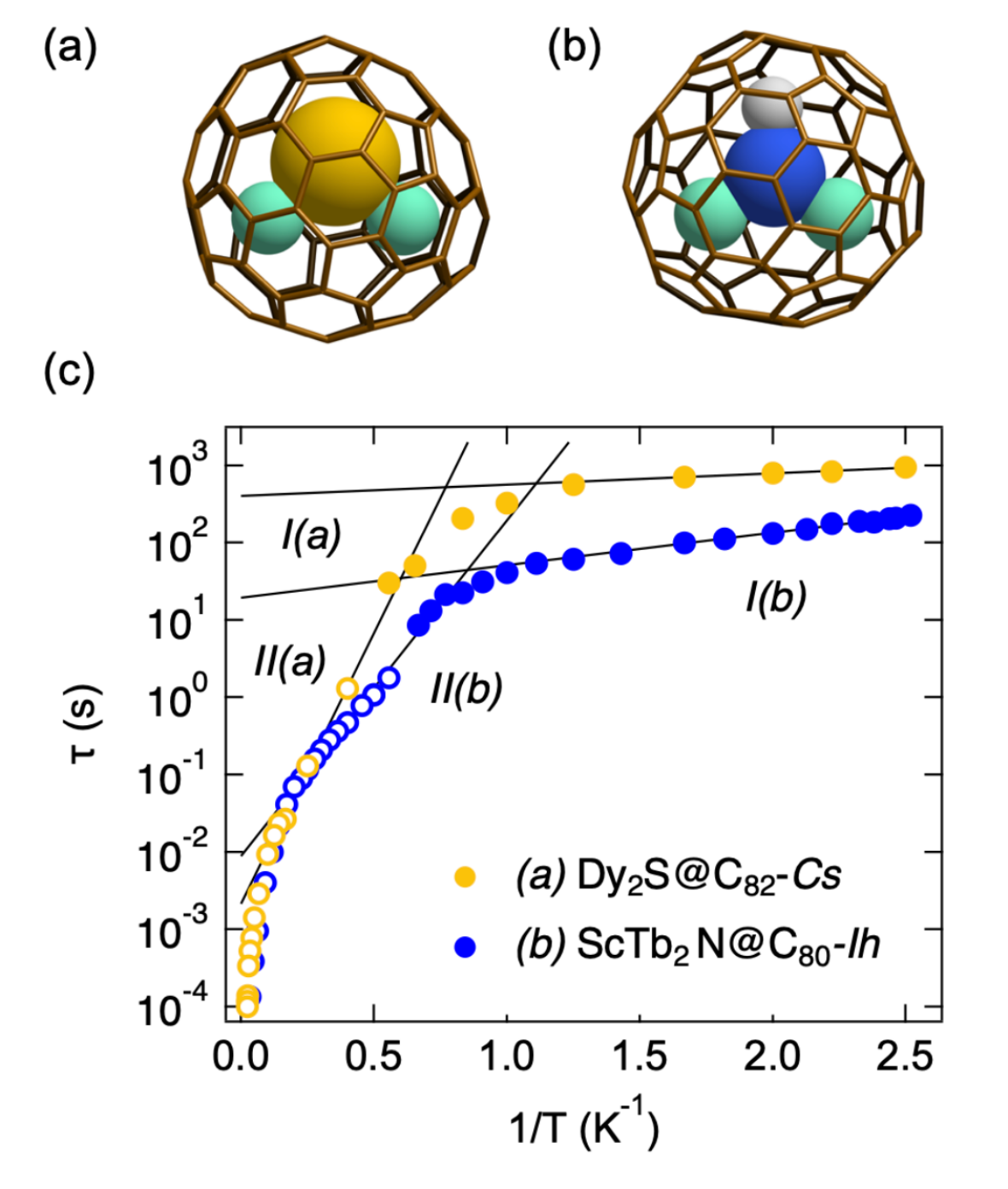}
    \caption {Dilanthanide single molecule magnets. (a) Model of \DSD-$C_s$. The sizes of the endohedral ions Dy (turquois) and S (yellow) are mimicked with their ion radii. (b) Model of \TTS-$I_h$. N (blue), Sc (grey), Tb (turquois). (c) Arrhenius plot of the zero field magnetisation lifetimes $\tau$ for \DSD-$C_s$ (yellow) \cite{Kry2020} and \TTS-$I_h$ (blue)  \cite{Kos2020}. Full symbols DC measurements, open symbols AC measurements. $I$ and $II$ are Arrhenius processes (see Table~\ref{T1}).
}
 \label{F1}
\end{figure}
\begin{table}[ht]
  \caption{Fit parameters of prefactors $\tau_{0}\,\mathrm{(s)}$ and barriers \Deff$/k_B \, \mathrm{(K)}$ of two Arrhenius processes for the temperature dependence of the zero field magnetization relaxation times $\tau$ for \DSD\ and \TTS\ in Figure~\ref{F1}(c).}
  \begin{center}
  \begin{tabular}{r|c|r|c|r}  

 \hline\hline
    \multirow{2}{*}{} &
      \multicolumn{2}{c|}{\DSD} &
      \multicolumn{2}{c}{\TTS } \\
    & $\tau_{0}$~(s) & \Deff$/k_B$ &  $\tau_{0}$~(s) & \Deff$/k_B$  \\ \hline 
    
   
    $I$ & $(4.0 \pm 0.3)\times 10^{~2}$& $0.34 \pm 0.03$ &$(1.9 \pm 0.2)\times 10^{~1}$&$0.97 \pm 0.04$  \\
    
  $II$ & $(2.1 \pm 1.3)\times 10^{{\text{-}}3} $    & $16.1 \pm 1.1$&$(8.9 \pm 1.0)\times 10^{{\text{-}}3} $&$10.0 \pm 0.2$  \\
  \hline\hline

  \end{tabular}
    \end{center}
\label{T1}
\end{table}
The term "pseudospin" is used because the ligand field splits the 2$J$+1 total angular momentum states, where for  $\{\mathrm{Tb, Dy, Ho}\}$ the $J_z{\text{=}}\{\pm6,\pm15/2, \pm8\}$ have the lowest energy. 
The excitation $\Delta$ to the second lowest $J_z$ level is large and in the order of several hundred $k_B$~K \cite{Zhang2015} which allows to neglect these excitations at low temperatures.

Figure~\ref{F1}(a) and (b) display the two endofullerenes \DSD-$C_s$  and \TTS-$I_h$ that serve as model systems for the investigation of the quantum tunneling of the magnetisation in axially anisotropic 4f ion pairs. 
The numbers of cage carbon atoms of 82 and 80 and the isomers $C_s$ and $I_h$ determine the stabilisation of (Dy$_2$S)$^{4+}$ \cite{Dunsch2010} and (Tb$_2$ScN)$^{6+}$ \cite{Nakao1994} endohedral units, respectively. As they do not affect the conclusions of this paper the cage isomer labels are omitted in the following. 
Within the picture developed for \DDS\ the zero field ground state splits in two time reversal symmetric doublets (TRD's)  \cite{Westerstrom2014}.The splitting is reflected in the temperature dependence of the zero field magnetisation lifetime. This is an empirical observation which implies that thermal fluctuations exceeding the energy splitting accelerate the reaching of the thermal equilibrium.
For \DSD\ the temperature dependence of the zero field magnetisation lifetime confirmed this picture and below 2~K it tended to level off, which was assigned to the onset of quantum tunneling of the magnetisation \cite{Kry2020}. This was not found for \TTS\, but a further Arrhenius barrier was identified down to 400~mK \cite{Kos2020}.
It was argued that this 1~K barrier is not explained within the ground state picture in Ref.~\cite{Westerstrom2014} and dipolar intermolecular interactions were offered as an explanation. As both molecules should have similar intermolecular magnetic interactions, but as no 1~K barrier was found in \DSD\ the intermolecular interaction hypothesis might not be correct. In the following we offer an intramolecular explanation that reveals the coherent pseudospin flip or tunneling rates in molecules with coupled anisotropic 4f ion pairs.
Figure~\ref{F1}(c) displays the published zero field magnetisation lifetimes of \DSD\ \cite{Kry2020} and \TTS\ \cite{Kos2020} in an Arrhenius plot in the temperature range between 0.4 and 30~K where the solid lines represent Arrhenius barriers $\tau_0\exp(\Delta_{{\rm{eff}}}/k_BT)$. The samples had natural isotope abundance.
In Table~\ref{T1} the fit-results of two distinct decay processes with barriers $\Delta_{{\rm{eff}}}^i$ and prefactors $\tau_{0,i}$ are listed for both molecules.
Process $II$ in the order of 10~K is the decay that is mediated via the TRD excitation \cite{Westerstrom2014}, and process $I$ is assigned to the quantum tunneling of the magnetisation that is shown here to cause a zero field splitting of the ground state.



\begin{figure}[b] 
\centering
\includegraphics[width=8 cm]{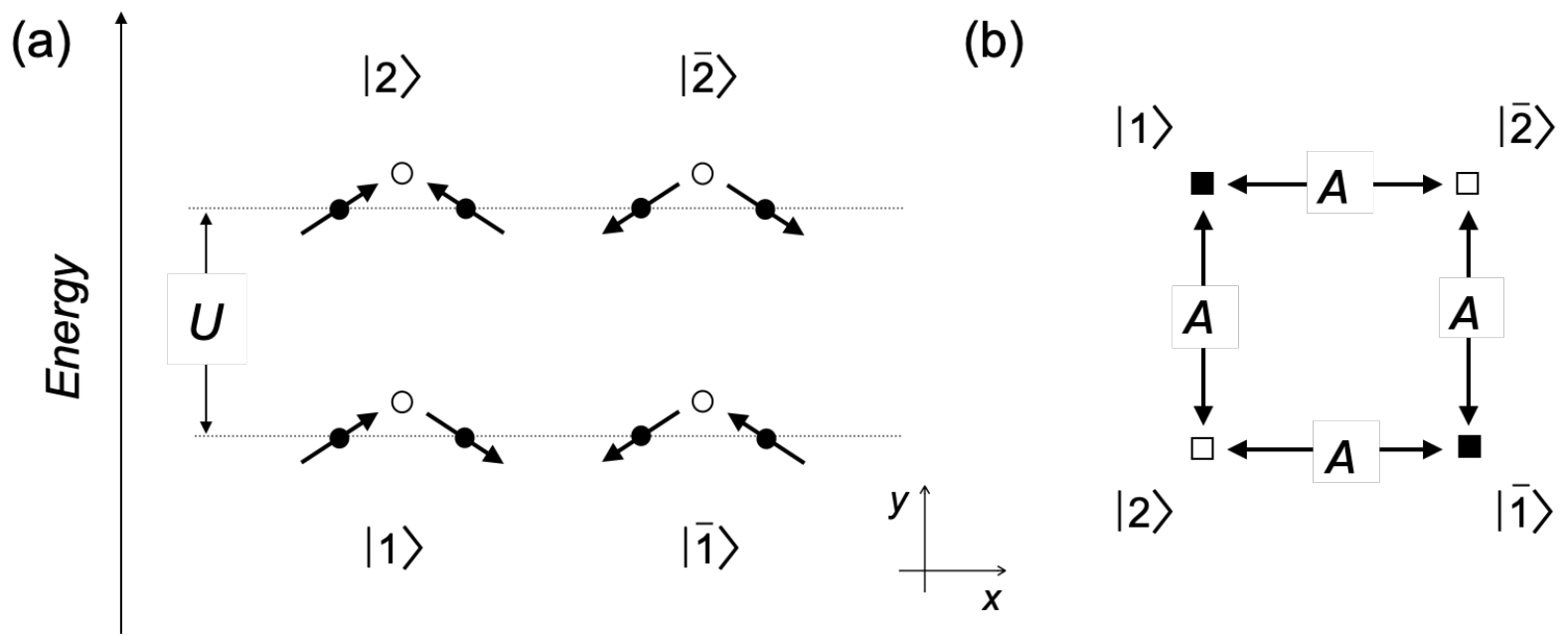}
\caption{Ground states. (a) Pseudospin configurations for axially anisotropic 4f ion pairs. The two pseudospins on the 4f ions (black) may point toward or away from the central anion (open circle). Four states $|1\rangle$, $|\bar{1}\rangle$, $|2\rangle$ and $|\bar{2}\rangle$ may form, where ($|1\rangle$, $|\bar{1}\rangle$) and ($|2\rangle$, $|\bar{2}\rangle$) are time reversal symmetric doublets (TRD's) with a dipolar and exchange splitting of~$U$. The resulting magnetic moments of TRD1 point along $\pm x$, and those of TRD2 along $\pm y$.
(b) The four states on a square in Hilbert space. Two adjacent states (corners) are separated by the flip of one pseudospin out of the two. $A$ is the tunneling matrix element for a pseudospin flip. The different energy (black or white) of adjacent states provided the rationale "exchange protection" for the large remanence of \DDS
\cite{Westerstrom2014}.}
\label{F2}
\end{figure}

Figure \ref{F2} sketches the model of the ground state of axially anisotropic 4f ion pairs like those in \DSD\ or \TTS\ in zero external magnetic field. 
The two pseudospins allow 2$^2$ possible ground state configurations that split by a dipolar and exchange interaction $U$ into two time reversal symmetric doublets TRD's ($|1\rangle$, $|{\bar{1}}\rangle$)  and ($|2\rangle$, $|{\bar{2}}\rangle$) \cite{Westerstrom2014}.
For the states ($|1\rangle$, $|{\bar{1}}\rangle$) the scalar product between the pseudospins is positive and they are called ferromagnetically coupled, while it is negative for the states ($|2\rangle$, $|{\bar{2}}\rangle$) that are antiferromagnetically coupled. While the angle between the two pseudospin axes affects the magnetisation curves \cite{Wes2021}, it is not of importance for the below conclusions on the quantum tunneling of the magnetisation. For \DDS , \TTS , \DSD\ and $\mathrm{Dy_2TiC@C_{80}}$ the ground states were found to be "ferromagnetic"  \cite{Westerstrom2014,Kos2020,Kry2020,Wes2021}, while \DDO\ has an antiferromagnetic groundstate \cite{Yang2019}. The below theory is applicable for both couplings. The energy difference $U$ can be extracted from the equilibrium magnetisation curves and the temperature dependence of the magnetisation lifetimes \cite{Westerstrom2014,Kos2020,Kry2020,Yang2019,Wes2021}.

The "Hilbert space topology" of the four states is depicted in Figure~\ref{F2}(b).
They lie on the corners of a square where the sides connect states that differ by one pseudospin flip.
It was anticipated that single pseudospin flips or tunneling within the four states was inhibited by the exchange and dipolar interaction $U$. The difference between \DDS\ and \TTS\ was assigned to the Kramers degeneracy of the odd 4f$^{\,9}$ Dy$^{3+}$ ions in contrast to the even 4f$^{\,8}$ Tb$^{3+}$ ions \cite{Kos2020,Kry2020}.


Without tunneling of the magnetisation the four states  $|{1}\rangle$,  $|{\bar{1}}\rangle$,  $|{2}\rangle$ and  $|{\bar{2}}\rangle$ are eigenstates of the system, where in zero field and with ferromagnetic coupling $|{1}\rangle$ and $|{\bar{1}}\rangle$ form a two fold degenerate ground state.
With  tunneling of the magnetisation the ground state degeneracy is lifted. 
This is seen if the Hamiltonian $\mathcal{H}$ for Figure~\ref{F2}(b) is written as a 4$\times$4 matrix (Equation~\ref{eq:H11}) and if the eigenvalues are determined. 
On the diagonal we find the energies as put forward in the picture of Westerstr\"om \cite{Westerstrom2014}.
Off diagonal we find the tunneling matrix element $A$ that describes the single spin flip probability. 
The sign of the tunneling matrix element is chosen such that the amplitudes of the ground state $|\Phi_1\rangle$ have the same sign.
\begin{equation}
\mathcal{H}=\begin{bmatrix}
    0 & 0 & \text{-}A & \text{-}A \\ 
    0 & 0 & \text{-}A & \text{-}A \\ 
    \text{-}A & \text{-}A & U & 0 \\ 
  \text{-}A & \text{-}A & 0 & U \\ 
    \end{bmatrix}
    \label{eq:H11}
    \end{equation}
   Here double flips $|1 \rangle \leftrightarrow |\bar{1} \rangle$ or $|2 \rangle \leftrightarrow |\bar{2} \rangle$ accross the diagonals in Figure~\ref{F2}(b) are neglected.
    An arbitrary state $|\Psi \rangle$ is described by 4 amplitudes $|a_{1},a_{\bar{1}},a_{2},a_{\bar{2}}\rangle$, where for example $|0,1,0,0\rangle$ corresponds to $|{\bar{1}}\rangle$, which is for $A\ne0$ not an eigenstate. The eigenstates are mixtures of the base in Figure~\ref{F2}. Importantly, the lifting of the zero field degeneracy of the ground state should enable pseudospin control and manipulation.
    The four eigenvalues and the eigenvector amplitudes for $U/A=10$ of  Eq.~\ref{eq:H11} are listed in Table~\ref{THH1}.
\begin{table}[b]
        \caption{Eigenvalues $\lambda_i$ and eigenvectors $|\Phi_i\rangle$ of the zero field Hamiltonian~\ref{eq:H11} with amplitudes $a_j$ in the basis of Figure~\ref{F2} for $U/A=10$ .}
  \begin{center}
  \begin{tabular}{ c | c |c|c|c|c  }
  \hline \hline
     Eigenvalue $\lambda_i$ &$|\Phi_i\rangle$ &  $a_1$&$a_{\bar{1}}$ &$a_{2}$&$a_{\bar{2}}$\\ \hline
 $\Big(U-\sqrt{16A^2+U^2}$ \Big)/2& $|\Phi_1\rangle$& 0.69&0.69&0.13&0.13\\ \hline
  0 &$|\Phi_2\rangle$&  -0.71&0.71&0&0\\ \hline
  $U$ &$|\Phi_3\rangle$&0&0&-0.71&0.71\\ \hline
  $\Big(U+\sqrt{16A^2+U^2}$ \Big)/2 &$|\Phi_4\rangle$& 0.13&0.13&-0.69&-0.69\\ \hline\hline
       \end{tabular}
      \end{center}

      \label{THH1}
\end{table}

\begin{figure}[h] 
\centering
\includegraphics[width=7 cm]{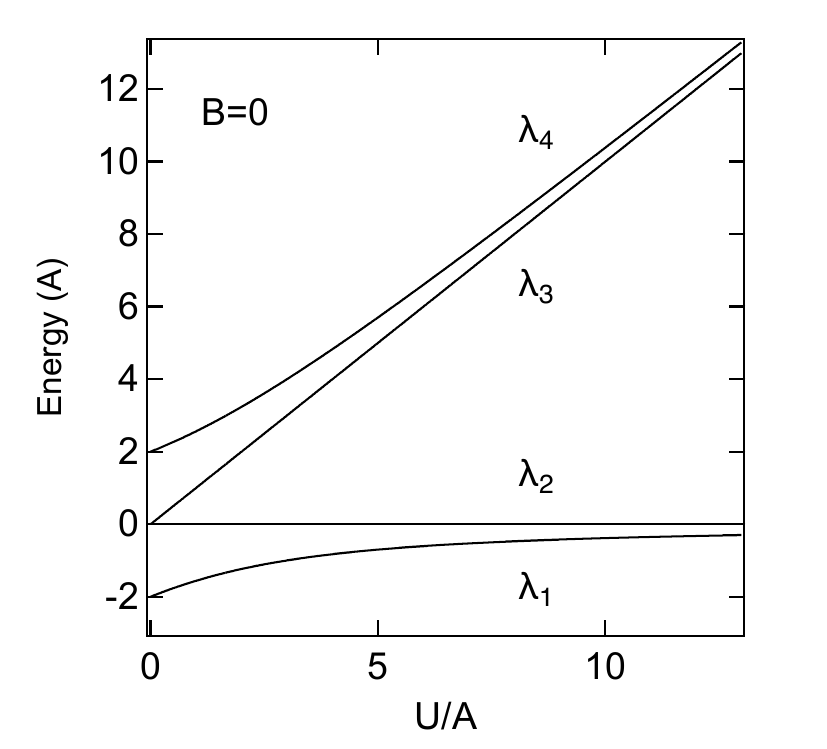}
\caption{Eigenvalues of the 4$\times$4 Hamiltonian (Eq.~\ref{eq:H11}) as a function of $U/A$ in zero field ($B$=0).} 
\label{FH2}
\end{figure}

Figure~\ref{FH2} shows the eigenvalue spectrum as a function of $U/A$.
Setting $(\lambda_2-\lambda_1)=$\Deff$^I$ we get for the two molecules frequencies \Deff$^I/h$ of 6.3 and 20.8~GHz, respectively.
For $U\gg A$ the splitting $(\lambda_2-\lambda_1)$ of the groundstate is $\approx 4A$.
Accordingly we get for the tunneling matrix element of \DSD\ and \TTS\ $A/k_B\text{=}$85 and 250~mK and $U/A\text{=}$40 and 190, respectively.
For Tb $A$ is in the range of the hyperfine interaction of isolated $^{159}$Tb$^{3+}$ with a nuclear spin $I\text{=}3/2$ \cite{Thiele2013}. 
The hyperfine interaction influences the magnetisation lifetimes of SMM's, though no strong influence on the quantum tunneling rate was found for isotope separated  Dy dimer SMM's \cite{Moreno2019}. It is therefore expected that the nuclear spin rather influences the prefactors $\tau_0$ but not the barriers in the $\ln \tau$ vs. $1/T$ data. 

\begin{figure}[b] 
\centering
\includegraphics[width=7 cm]{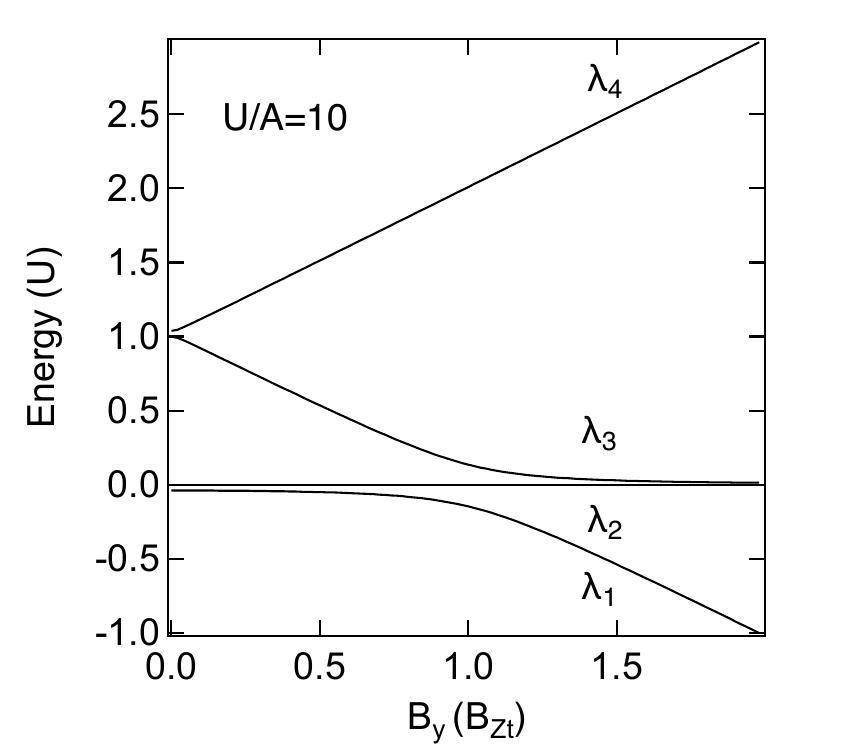}
\caption{Eigenvalues $\lambda_i$ of ~\ref{eq:HH2}  as a function of an external magnetic field along $y$ for $U/A=10$. The B-field scale is $B_{Zt}=U/2\mu_y$.}
\label{FH222}
\end{figure}

The magnetic moment of the states can be calculated from $\langle \Psi|M|\Psi\rangle$, where $M=(\vec{\mu}_1,\text{-}\vec{\mu}_1,\vec{\mu}_2,\text{-}\vec{\mu}_2)$ are the magnetic moments of the base. It is seen that all four eigenstates of \ref{eq:H11} have no magnetic moment.


In an external B-field this changes. The Hamiltonian~\ref{eq:H11} has to be complemented with corresponding Zeeman terms {\mbox{$E^Z_{j}=- \boldsymbol{\mu}_j\cdot \bf{B}$}}, where $\boldsymbol{\mu}_j=(\pm 2\mu_x,\pm 2\mu_y)$ are the magnetic moments of one of the four pseudospin configurations, and ${\bf B}$ the magnetic field.
For the state conventions in Figure~\ref{F2} we get for a B-field $B_y$ and a pseudospin magnetic moment $\mu_y$ along~$y$, $E^Z_1=0$ and $E^Z_2=2\mu_y B_y$ and  $\mathcal{H}$ in \ref{eq:H11} extends to $\mathcal{H}^Z_y$:
\begin{equation}
\mathcal{H}^Z_y=\begin{bmatrix}
    0 & ~~0 &  \text{-}A &  \text{-}A \\ 
    0 & ~~0 & \text{-}A & \text{-}A \\ 
     \text{-}A &  ~\text{-}A & ~U\text{-}E^Z_2 & 0 \\ 
     \text{-}A &  ~\text{-}A & 0 & ~U\text{+}E^Z_2 \\ 
    \end{bmatrix}
    \label{eq:HH2}
    \end{equation}

The effect of the hybridisation displays in a plot of the eigenvalues vs. the applied B-field
(Figure~\ref{FH222}), where the field scale is chosen to be the Zeeman threshold field $B_{Zt}\equiv U/2\mu_y$ \cite{Kos2020}, which is 1.9 and 1.6~T for \DSD\ and \TTS, respectively. 
The level crossing at $B_{Zt}$ is described by the Hamiltonian~\ref{eq:HH2}, which is of help for the understanding of kinks in hystereses.
The eigenvector of the ground state $|\Phi_1\rangle$ for B=0 without magnetic moment evolves for large fields non-linearly to state $|2\rangle$ with a magnetisation along $y$, while the first excited state $|\Phi_2\rangle$ remains without magnetic moment and constant in energy.   
Of course, the external {\mbox{B-field}} can be applied in an arbitrary orientation, though for the field along $y$ $\lambda_2$ remains constant, which can be of particular interest.

In conclusion, the time dependence of the magnetisation of the two endofullerene single molecule magnets \DSD\ and \TTS\ at sub-Kelvin temperatures is revisited and used to extract quantum tunneling rates of the magnetisation which lie in the GHz range. The proposed model completes the description of the ground state, is robust and can be applied to any coupled anisotropic 4f ion pair. While all states have no magnetic moment in zero field the ground state displays a non-linear magnetic moment in weak applied fields. These findings are expected to facilitate the search for quantum behaviour in such systems where coherent tunneling of the magnetisation shall be exploited.  

Financial support from the Swiss National Science Foundation (SNF project 201086) is gratefully acknowledged. \mbox{Ari P. Seitsonen} designed Figure~\ref{F1}(a) and (b).

%

\end{document}